\documentclass[aps,prd,preprint,superscriptaddress,preprintnumbers,nofootinbib]{revtex4-1}
\pdfoutput=1

\usepackage{graphics}
\usepackage[dvips]{graphicx}
\usepackage{mathrsfs}
\usepackage{amssymb}
\usepackage{amsmath}
\usepackage{verbatim}
\usepackage{float}
\usepackage{slashed}
\usepackage{bbm}
\usepackage{subfigure}
\usepackage[dvips,letterpaper,text={6.5in,9in}]{geometry}

\usepackage[english]{babel}
\usepackage{mwe}

\usepackage{latexsym}
\usepackage{mathrsfs}
\usepackage{amsthm}
\usepackage{tabu, multirow,color,dcolumn,bm,enumerate}
\newcommand{\tev}{\text{TeV}}
\newcommand{\gev}{\text{GeV}}

\usepackage{graphicx,bm}

\newcommand{\beq}{\begin{equation}}
\newcommand{\bea}{\begin{eqnarray}}
\newcommand{\eeq}{\end{equation}}
\newcommand{\eea}{\end{eqnarray}}
\newcommand{\bal}{\begin{align}}
\newcommand{\eal}{\end{align}}

\newcommand{\be}{\begin{equation}} 
\newcommand{\ee}{\end{equation}} 
\newcommand{\bs}{\begin{split}} 
\newcommand{\es}{\end{split}}

\newcommand{\tr}{\operatorname{tr}}

\newcommand{\Lc}{\mathcal{L}}

\begin{document}

\title{Dark matter from the supersymmetric  custodial triplet model}

\author{Antonio Delgado}
\affiliation{Department of Physics, 225 Nieuwland Science Hall, University of Notre Dame, Notre Dame, IN 46556, USA}
\affiliation{Theory Division, Physics Department, CERN, CH-1211 Geneva 23, Switzerland}
\author{Mateo Garcia-Pepin}
\affiliation{Institut de F\'isica d'Altes Energies, Universitat Aut{\`o}noma de Barcelona, 08193 Bellaterra, Barcelona, Spain}
\author{Bryan Ostdiek}
\affiliation{Department of Physics, 225 Nieuwland Science Hall, University of Notre Dame, Notre Dame, IN 46556, USA}
\email[E-mail: ]{bostdiek@nd.edu}
\author{Mariano Quiros}    
\affiliation{Institut de F\'isica d'Altes Energies, Universitat Aut{\`o}noma de Barcelona, 08193 Bellaterra, Barcelona, Spain}
\affiliation{Instituci{\'o} Catalana de Recerca i Estudis  
Avan\c{c}ats (ICREA) and Institut de F\'isica d'Altes Energies, Universitat Aut{\`o}noma de Barcelona,
08193 Bellaterra, Barcelona, Spain}

\preprint{CERN-PH-TH/2015-078} 

\begin{abstract}
\vspace*{0.5cm}
The supersymmetric custodial triplet model adds to the particle content of the MSSM three $SU(2)_L$ triplet chiral superfields with hypercharge $Y=(0,\pm1)$. At the superpotential level the model respects a global $SU(2)_L \otimes SU(2)_R$ symmetry only broken by the Yukawa interactions. The pattern of vacuum expectation values of the neutral doublet and triplet scalar fields depends on the symmetry pattern of the Higgs soft breaking masses. We study the cases in which this symmetry is maintained in the Higgs sector, and in which it is broken only by the two doublets attaining different vacuum expectation values. In the former case, the symmetry is spontaneously broken down to the vectorial subgroup $SU(2)_V$ and the $\rho$ parameter is protected by the custodial symmetry. However in both situations the $\rho$ parameter is protected at tree level, allowing for light triplet scalars with large vacuum expectation values. We find that over a large range of parameter space, a light neutralino can supply the correct relic abundance of dark matter either through resonant s-channel triplet scalar funnels or well tempering of the Bino with the triplet fermions. Direct detection experiments have trouble probing these model points because the custodial symmetry suppresses the coupling of the neutralino and the $Z$ and a small Higgsino component of the neutralino suppresses the coupling with the Higgs. Likewise the annihilation cross sections for indirect detection lie below the Fermi-LAT upper bounds for the different channels. 
\end{abstract}

\maketitle

%*********************Introduction***********************	
\section{Introduction}
\label{sec:intro}
%*********************Introduction***********************

Although the Standard Model (SM) of strong and electroweak interactions is incredibly successful in explaining all high- and low-energy particle physics data, it is known to be incomplete. One such reason is the astrophysical observation of dark matter leading to the belief that there should exist a particle explaining it.  However there is no candidate dark matter particle in the Standard Model. Moreover cosmic microwave background measurements can be fitted extremely well with  a cosmological  $\Lambda \text{CDM}$ model if the relic density of dark matter is given by $\Omega_{\text{DM}} h^2 = 0.1187$~\cite{Ade:2013zuv, Planck:2015xua}.

On the other hand, the recent discovery of the Higgs by the ATLAS~\cite{Aad:2012tfa} and CMS~\cite{Chatrchyan:2012ufa} experiments seems to point toward a single $SU(2)_L$ doublet as being responsible for the breaking of the electroweak symmetry. However, this cannot be known for sure without more precise measurements of its properties. In fact, beyond the Standard Model proposals predict deviations in the couplings of the Higgs compared to the SM values and can alleviate any possible future discrepancies between the predicted and observed properties. The simplest such models are those with extended Higgs sectors. A drawback for extended Higgs sectors is that they can run into trouble with the $\rho$ parameter if the extra $SU(2)_L$ representations are large enough.\footnote{Extra doublets (and singlets) do not suffer from this problem. However representations beyond that do.} Models preserving custodial symmetry, for instance, a septet under $SU(2)_L$ with hypercharge $Y=2$~\cite{Kanemura:2013mc, Hisano:2013sn, Kanemura:2014bqa, Killick:2013mya, Alvarado:2014jva, Geng:2014oea} or the Georgi-Machacek (GM) model, which contains a real and a complex scalar triplet with $Y=(0,1)$, respectively~\cite{Georgi:1985nv, Chanowitz:1985ug, Gunion:1989ci, Gunion:1990dt, Gunion:1989we, Haber:1999zh, Aoki:2007ah, Godfrey:2010qb, Low:2010jp, Low:2012rj, Logan:2010en, Falkowski:2012vh, Chang:2012gn, Chiang:2012cn, Kanemura:2013mc, Englert:2013zpa, Killick:2013mya, Englert:2013wga, Chiang:2013rua, Efrati:2014uta, Hartling:2014zca, Chiang:2014hia, Chiang:2014bia, Godunov:2014waa, Hartling:2014xma,Hartling:2014aga, Chiang:2015kka,Logan:2015xpa}, can solve this problem and keep $\rho=1$ at tree level. However, there is nothing in the previous models protecting the scalar masses from large radiative corrections, so all of them suffer from a more severe hierarchy problem than the Standard Model.

Supersymmetry provides a nice solution to both the dark matter problem and the hierarchy problem. If R parity is assumed, the lightest supersymmetric particle (LSP) is stable. For the model to be cosmologically viable, the LSP must be neutral, thus providing a dark matter candidate. However, not all dark matter candidates yield the observed relic abundance. Only specific regions of parameter space will allow the lightest neutralino to freeze out to the observed relic abundance. In addition to this, the minimal supersymmetric Standard Model (MSSM) runs into difficulties when trying to fit the observed Higgs mass. One method of raising the Higgs mass in supersymmetric models is to extend the Higgs sector beyond that of the MSSM. The supersymmetric custodial triplet model (SCTM)~\cite{Cort:2013foa, Garcia-Pepin:2014yfa} (a supersymmetric generalization of the GM model) does this by introducing new $F$-term contributions to the tree-level MSSM Higgs mass. Besides, a custodial potential is interesting from a dark matter perspective. The coupling of the $Z$ to the neutralinos vanishes at tree level in the custodial limit of the MSSM (for $\tan\beta=1$), leading to blind spots in the spin-dependent dark matter searches~\cite{Cheung:2012qy}. A custodially symmetric extended Higgs sector will maintain this property.

It was argued in Ref.~\cite{Garcia-Pepin:2014yfa} that a totally custodial situation at the electroweak (EW) scale is not favored from a theoretical point of view. The authors examined this issue by imposing a global $SU(2)_L\otimes SU(2)_R$ symmetry at a high scale, which would then be broken through the renormalization group running of parameters down to the EW scale. Because of the influence of the top quark Yukawa coupling the running differentiates the two soft doublet masses from each other much more than the three triplet ones among themselves, resulting in a much bigger vacuum misalignment in the doublet sector. The difference in the doublet sector results in a departure from $\tan\beta=1$.  As the $\rho$ parameter is affected only by the difference in the triplet VEVs, the model still allows the triplets to contribute up to $15\%$ to the breaking of the EW symmetry.

In our study, we take the middle ground between the calculable fully custodial model of Ref.~\cite{Cort:2013foa} and the model of Ref.~\cite{Garcia-Pepin:2014yfa} in which an ultraviolet completion is proposed. We assume a Higgs sector with a potential allowing for a noncustodial vacuum, provided that this only comes from the ratio of the doublet vacuum expectation values (VEVs), parameterized by $\tan\beta$. This turns out to be a very good approximation to the situation explored in Ref.~\cite{Garcia-Pepin:2014yfa}. Within this approach, we explore the dark matter properties of the model and find that there are large regions of parameter space where dark matter annihilation in the early Universe occurs through the new triplet states. We also examine the direct detection consequences of breaking the custodial symmetry along the $\tan\beta$ direction and the indirect detection bounds.

The rest of the paper is organized as follows. In Sec.~\ref{sec:NeutralinoDM}, we briefly review the conditions needed to generate the correct relic abundance of neutralino dark matter. Section~\ref{sec:model} introduces the model and the benchmark parameters used for our study. The scalar spectrum is studied in Sec.~\ref{sec:Scalar}. The mixing of the neutralinos is discussed in Sec.~\ref{sec:NeutralinoMixing} which leads into the study of the relic abundance of dark matter and direct detection constraints of the model in Sec.~\ref{sec:darkmatter}. We discuss our conclusions in Sec.~\ref{sec:Discussion}.

%*********************NeutralinoDM***********************
\section{Neutralino dark matter}
\label{sec:NeutralinoDM}
%*********************NeutralinoDM***********************

In this paper, we focus on dark matter coming from the neutralino sector of the SCTM~\cite{Cort:2013foa, Garcia-Pepin:2014yfa} in which the content of the MSSM is extended by three triplets with hypercharge $Y=(0,\pm1)$. This adds three neutralinos to the four MSSM ones, the Bino, Wino, and the two Higgsinos. There are also two new charginos on top of the charged Wino and Higgsino, and finally a doubly charged triplet fermion. We will collectively refer to the fermion components of all of the triplet fields as tripletinos. The combination of the neutralinos, charginos, and doubly charged tripletinos will be referred to as electroweakinos.

If the mass parameters of the electroweakinos are well separated, mixing can be neglected and the LSP can be a pure gauge eigenstate. The pure Bino does not annihilate enough in the early Universe, while both the Wino and Higgsino annihilate easily and need a mass near or above a TeV in order to freeze out with the correct relic abundance. If their masses are lighter than this, the pure Wino or Higgsino leaves too little dark matter. The pure Wino may already be excluded by astrophysical gamma ray searches.\footnote{As $SU(2)_L$ triplets, the tripletinos should behave similarly to the Wino in this regard.} After constraints from LEP, the LHC, and astrophysics are applied, the only pure state that can generate the observed relic abundance is the Higgsino~\cite{lepii, Cohen:2013ama,Fan:2013faa}.

To have neutralino dark matter lighter than a TeV and freeze out to the observed relic abundance, the LSP must have a large Bino component, and there must be a process which helps the LSP to annihilate efficiently in the early universe. There are a few options to increase the rate at which the Bino annihilates. 
\begin{enumerate}
	\item Mixing: If the composition of the LSP contains a substantial amount of Wino, Higgsino or Tripletino, the mixing can allow for efficient annihilations.
	\item Coannihilation: Having another supersymmetric particle slightly above the mass of the LSP opens the possibility of $t$-channel annihilations, which can greatly increase the annihilation cross section. For there to be enough of the heavier particle around as the universe expands and cools down, the mass must not be more than $\sim10\%$ larger than the mass of the dark matter (DM) candidate.
	\item Funnel/Resonance: If the mass of the LSP is approximately half the mass of another state, the $s$-channel propagator becomes very large. There is a peak in the annihilation cross section, and a corresponding dip in the relic abundance after freeze-out.
\end{enumerate}

If the LSP is coannihilating with squarks or sleptons, there are strong limits on the model from LHC searches. This is due to the production rate of squarks and the relatively clean signals for sleptons. In this case, one would expect to find the squark or slepton before the dark matter candidate. 

In the literature, both mixing and coannihilation among electroweakinos are referred to as well tempering~\cite{ArkaniHamed:2006mb}.  Well tempering implies that there are multiple states around which can be produced, which is good for the production cross section of beyond-the-Standard-Model states. However, achieving the correct relic abundance requires the splitting to be small, which makes detection difficult. There have recently been studies on detecting electroweakinos with small splittings at colliders~\cite{Baer:2009bu, Schwaller:2013baa,Low:2014cba, Bramante:2014dza, Calibbi:2014lga, Han:2014sya, Bramante:2014tba, Martin:2014qra, Han:2014xoa}. 

The resonant/funnel annihilations of the LSP do not need extra particles at the same mass, but instead at nearly twice the mass of the dark matter particle. In the MSSM, the funnel particle can be either of the {\it CP}-even Higgs, ($H^0_1, H^0_2$) or {\it CP}-odd Higgs ($A^0$)~\cite{Ellis:2012aa,Fowlie:2012im, Cohen:2013kna, Hooper:2013qjx, Han:2013gba, Henrot-Versille:2013yma, Anandakrishnan:2014fia}. The dark matter particle itself cannot be detected at colliders, which implies the way to look for such a model is through the heavier states. However, searches for neutral scalars are difficult, as exemplified by the long search for the Higgs. As will be shown later, in the SCTM, the triplet scalars provide a resonant channel over much of the parameter space. Because of to the degeneracy of states in the custodial situation~\cite{Cort:2013foa}, there are charged states near the neutral funnel that could aid in discovery. 

%*********************The Model***********************	
\section{Model}
\label{sec:model}
%*********************The Model***********************

As in Ref.~\cite{Cort:2013foa}, we will construct the supersymmetric Higgs sector manifestly invariant under $SU(2)_L\otimes SU(2)_R$. The MSSM Higgs sector $H_1$ and $H_2$ with respective hypercharges $Y=(-1/2,\,1/2)$ ,
 \be
   H_1=\left( \begin{array}{c}H_1^0\\ H_1^-\end{array}\right),\quad
   H_2=\left( \begin{array}{c}H_2^+\\ H_2^0\end{array}\right)
   \ee
 is complemented with $SU(2)_L$ triplets, $\Sigma_{Y}$, with hypercharges  $Y=(-1,\, 0,\, 1)$ 
 \be
 \Sigma_{-1}=\left(\begin{array}{cc} \frac{\chi^-}{\sqrt{2}} & \chi^0\\\chi^{--}& -\frac{\chi^-}{\sqrt{2}}
 \end{array}
 \right),\quad  \Sigma_{0}=\left(\begin{array}{cc} \frac{\phi^0}{\sqrt{2}} & \phi^+\\ \phi^{-}& -\frac{\phi^0}{\sqrt{2}}
 \end{array}
 \right),\quad  \Sigma_{1}=\left(\begin{array}{cc} \frac{\psi^+}{\sqrt{2}} & \psi^{++}\\\psi^{0}& -\frac{\psi^+}{\sqrt{2}}
 \end{array}
 \right)\ .
 \ee
where $Q=T_{3L}+Y$.

The two doublets and the three triplets are organized under $SU(2)_L\otimes SU(2)_R$ as $\bar H=(\textbf{2},\bar {\textbf{2}})$, and $\bar \Delta=(\textbf{3},\bar{\textbf{3}})$ where
\be
\bar H=\left( \begin{array}{c}H_1\\ H_2\end{array}\right),\quad
\bar\Delta=\left(\begin{array}{cc} -\frac{\Sigma_0}{\sqrt{2}} & -\Sigma_{-1}\\ -\Sigma_{1}& \frac{\Sigma_0}{\sqrt{2}}\end{array}\right)
\ee
and $T_{3R}=Y$.   The invariant products for doublets $A\cdot B\equiv A^a\epsilon_{ab}B^b$  and antidoublets $\bar A\cdot \bar B\equiv\bar A_a\epsilon^{ab}\bar B_c$ are defined by $\epsilon_{21}=\epsilon^{12}=1$. \\
The $SU(2)_L\otimes SU(2)_R$ invariant superpotential is defined as
\be
W_0=\lambda \bar H\cdot \bar\Delta\bar H+\frac{\lambda_3}{3}\tr\bar\Delta^3+\frac{\mu}{2}\bar H\cdot\bar H+\frac{\mu_\Delta}{2}\tr \bar\Delta^2
\label{W0}
\ee
and the total potential
\be
V=V_F+V_D+V_{\rm soft}
\ee
where

\begin{eqnarray}
V_{\rm soft}&=&m_{H_1}^2|H_1|^2+m_{H_2}^2|H_2|^2+m_{\Sigma_1}^2 \tr |\Sigma_1|^2+m_{\Sigma_{-1}}^2 \tr |\Sigma_{-1}|^2+m_{\Sigma_0}^2 \tr |\Sigma_0|^2
\nonumber\\
&+&\left\{\frac{1}{2}m_3^2\bar H\cdot\bar H
+ \frac{1}{2}B_\Delta\tr\bar\Delta^2+A_\lambda \bar H\cdot \bar\Delta \bar H+\frac{1}{3}A_{\lambda_3}\tr\bar\Delta^3+h.c.\right\}
\label{Vsoft}
\end{eqnarray}

Note that the potential we just wrote is the same as in Ref~.\cite{Cort:2013foa} but with noncustodial soft masses. They will be used to satisfy the equations of motion. The neutral components of all the fields can be parametrized by
\be
X = \frac{1}{\sqrt{2}}\left(v_X + X_R +\imath X_I \right), ~~~X=H^0_1,H^0_2,\phi^0,\chi^0,\psi^0.
\ee
By imposing
\be
\label{vacuum}
 v_1=\sqrt{2}\cos{\beta}v_H,\quad v_2=\sqrt{2}\sin{\beta}v_H\quad \textrm{and}\quad v_\psi=v_\chi=v_\phi\equiv v_\Delta,
\ee 
the custodial symmetry is only broken in the vacuum by $\tan\beta$. For the rest of the paper we refer to $\tan\beta=1$ as the custodial case and $\tan\beta\ne1$ as the noncustodial case. To set the $Z$ mass, the total VEV must be
\be
\label{vev246}
v^2 \equiv (246~\gev)^2= 2v_H^2+8v_{\Delta}^2.
\ee
The $\rho$ parameter is not affected if custodial symmetry is broken in this way. This will generate five equations of minimum that we will use to solve for the values of the soft masses. The minimization conditions can be found in Appendix \ref{sec:MinCon}. While Eq.~\eqref{condiciones} (and the previous noncustodial equations) provide the necessary conditions for an extremum of the vacuum, they do not guarantee a minimum. To guarantee the appearance of a minimum, it is a sufficient condition that the determinant of the Hessian at the origin be negative. In the custodial case, to leading order in small $v_{\Delta}$, this condition can be expressed as
\be
\lambda(2\mu-\mu_{\Delta})-A_{\lambda} > 0~~\text{ and }~~ \frac{3}{2} v_H^2 \lambda^2 -2m_3^2 <0.
\label{eqn:custmincond}
\ee
The left equation sets the allowed relative size between the doublet and triplet supersymmetric masses. The right equation will be very important as we scan across the parameter space. For a fixed value of $m_3^2$, the right equation sets a maximum value for $v_H^2\lambda^2$. The $\lambda$ parameter will be used to raise the tree-level Higgs mass, so there exist regions where the Higgs mass cannot be achieved with light stops while keeping the potential correctly minimized (i.e. not getting tachyonic states).

To begin a study of the dark matter properties of the model, we first choose a set of benchmark values, given by
\be
\begin{aligned}
\lambda_3&=0.35,\\
m_3 &= 500~\gev, \\
B_{\Delta}&=-(500 ~\gev)^2,\\
A_{\lambda}&=A_{\lambda_3}=A_t=A_b=A_{\tau}=0, \\
m_{\tilde{Q}_3}&= 800~\gev, ~~\text{and}~~ m_{\tilde{u}_3^c} = 700~\gev,
\label{eqn:benchmark}
\end{aligned}
\ee
where other scalar soft masses have been decoupled and the ones corresponding to Higgs multiplets are determined by the minimization conditions. The SCTM triplet $F$ terms yield a large tree-level Higgs mass, so smaller one-loop corrections are needed. This is the reason for our choice of relatively light stops albeit above the current experimental limits. The value of $\lambda_3$ will not have much of an effect. We are considering the case $m_3^2=|B_\Delta|$ for simplicity. Values of $m_3$ and $B_\Delta$ around those in Eq.~(\ref{eqn:benchmark}) should provide similar results. Larger values for $m_3$ or $B_{\Delta}$ will decouple the heavy scalars more, and in addition will affect how large $v_{\Delta}$ can be in the minimization of the potential. Similarly, we choose to examine the case in which all of the trilinear terms are zero to help ensure that the EW vacuum is the deepest one. This leaves $\mu$, $\mu_{\Sigma}$, $\lambda$, and $v_{\Delta}$ as the remaining free parameters to study.

%*********************ScalarMasses***********************
\section{Scalar masses}
\label{sec:Scalar}
%*********************ScalarMasses***********************
There is a total of five {\it CP}-even, five {\it CP}-odd, six singly charged, and two doubly charged Higgs scalar fields in this model. The mass matrices for all of these states are cumbersome, and not entirely enlightening. In Ref.~\cite{Cort:2013foa}, the fields were decomposed into the $SU(2)_V$ custodial basis, which is also directly related to the mass basis for $\tan\beta=1$, up to small hypercharge breaking effects. This notation can be helpful for showing how many charged states will be in the proximity of a neutral state. However, since we will be examining both the custodial and the noncustodial setups of the model, we will not use this notation. Instead, we will work with mass eigenstates. After removing the Goldstone bosons, they will be denoted as $H^0_{1,\dots 5}$, $A^0_{1,\dots 4}$, $H^+_{1,\dots 5}$, and $T^{++}_{1,2}$.

To study the dark matter annihilation in the model, we are really only interested in the spectrum of the lightest neutral scalars rather than the charged components. Annihilating the neutralino through a resonance of the Higgs or the heavy Higgs has been shown before in the MSSM. As a new feature of this model, there are substantial regions of parameter space in the SCTM where the annihilation can proceed through a tripletlike resonance. To do this, the soft masses of the triplets must not be too large. Upon close examination of the minimization conditions for $m^2_{\Sigma_0}$, $m^2_{\Sigma_1}$, and $m^2_{\Sigma_{-1}}$ in Eqs.~\eqref{eqn:minSig0}, \eqref{eqn:minSig1}, and \eqref{eqn:minSigm1} respectively, we see that there is a piece that scales as $v^2_H/v_{\Delta}$ for each soft mass. Smaller values for $v_{\Delta}$ yield large soft masses for the triplets, decreasing the chance of annihilating through the triplet funnel.

In the decoupling limit of the MSSM, the tree-level Higgs mass goes as $m^2_h = m_Z^2 \cos^2(2\beta).$ Because of this, for $\tan\beta=1$, there is no tree-level contribution to the mass of the Higgs from the MSSM parts of our model. Instead, the mass at tree level in the decoupling limit comes only from the triplet $F$ terms, and is given (at leading order in $v_\Delta$) by
\be
\left.m^2_h\right|_{\tan\beta=1} = 3 \lambda^2 v_H^2.
\ee
We also examine the model in which $\tan\beta\ne1$ but is still small. The tree-level Higgs mass can no longer be written in a simple form. However, we comment that there are now MSSM contributions to the mass, and the triplet $F$ terms contribute as $\lambda^2 \left(4 \cos^4 \beta + 4 \sin^4\beta + \sin^22\beta \right)$. The SCTM allows for large tree-level contributions to the Higgs mass with no need of large one-loop corrections, and thus no need for heavy stops.

The dominant radiative corrections to the Higgs mass depend on the top Yukawa coupling, defined as
\be
h_t = \frac{m_t}{v_2/\sqrt{2}}=\frac{m_t}{\sin\beta ~v_H} = \frac{m_t}{\sin\beta \sqrt{\left(v^2- 8v_{\Delta}^2\right)/2}}
\ee
They have been proven to be sizeable in the context of the MSSM, without jeopardizing perturbativity, as the top Yukawa coupling does not enter the Higgs mass at the tree level. In fact in the SCTM, increasing $v_{\Delta}$ increases the top Yukawa, which increases radiative corrections to the Higgs mass. In our study, we take the dominant one-loop corrections found in Ref.~\cite{Carena:1995bx},~\footnote{We will neglect radiative corrections proportional to $\lambda^2$ as the parameter $\lambda$ affects the Higgs mass at the tree level and thus the corresponding radiative corrections are constrained to be small by perturbativity.} we use $700~\gev$ for the right-handed soft mass and $800~\gev$ for the left-handed soft mass. These were chosen to be slightly above the current experimental bounds, regardless of the mass of the lightest neutralino. Raising the masses of the stops will not affect our dark matter results, only worsen the fine-tuning of the model. Note that even though the stop masses and $A_t$ are fixed in the study, changing $\mu$ and $\tan\beta$ affects the mixing and thus the one-loop contributions to the Higgs mass. 
\begin{figure}[h]
\begin{center}
\includegraphics[width=0.38\textwidth]{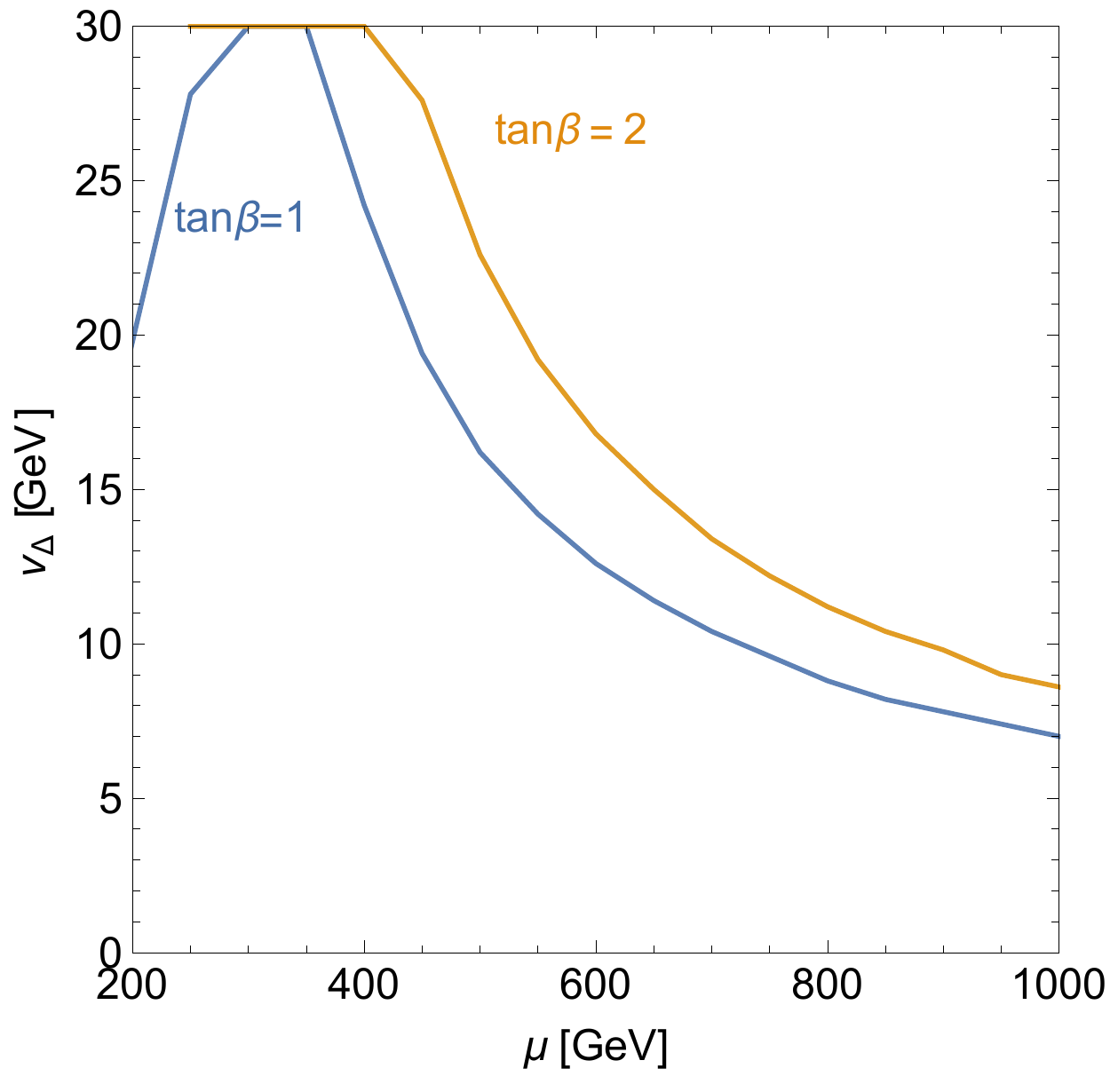}
\includegraphics[width=0.39\textwidth]{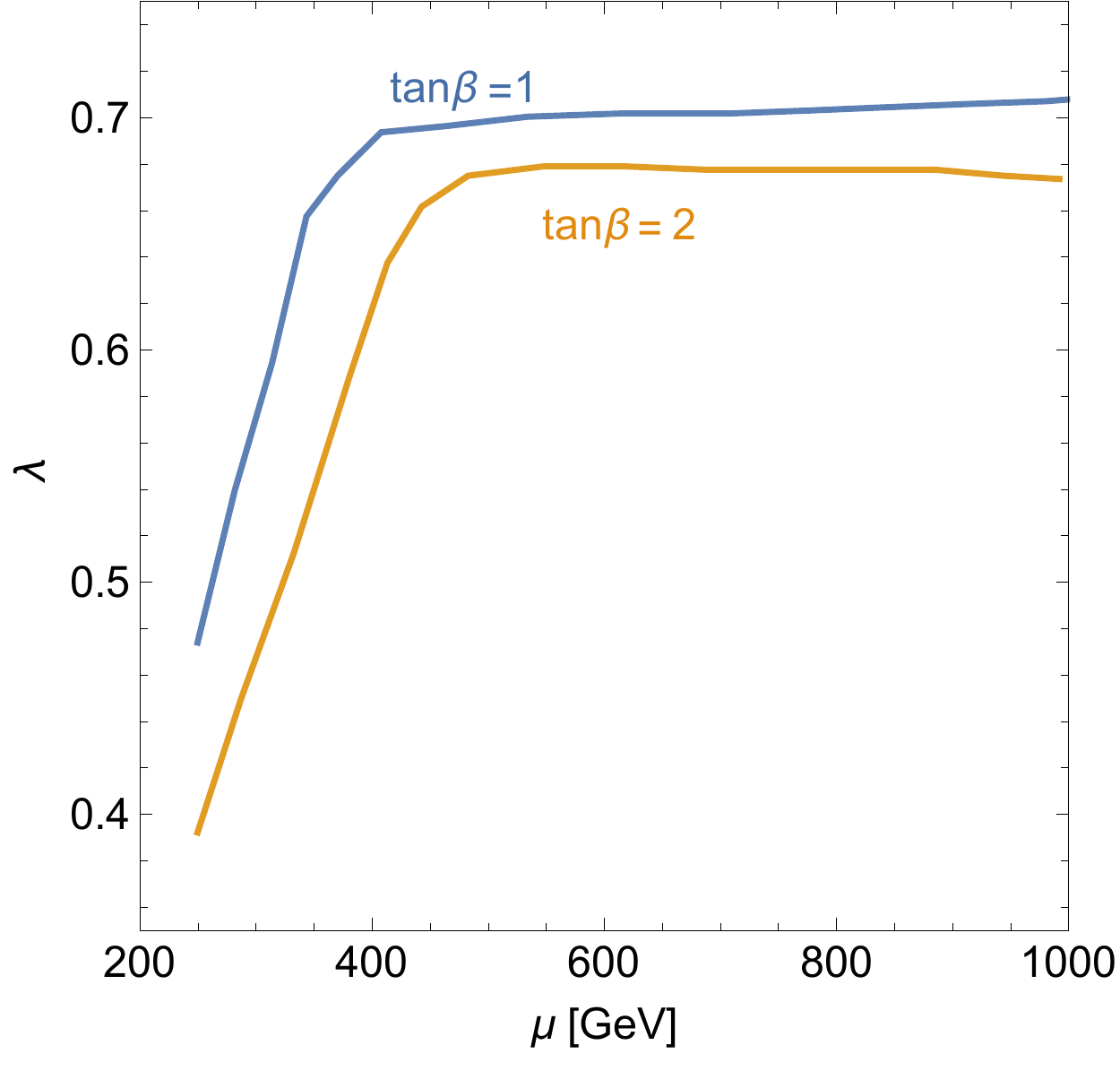}
\includegraphics[width=0.38\textwidth]{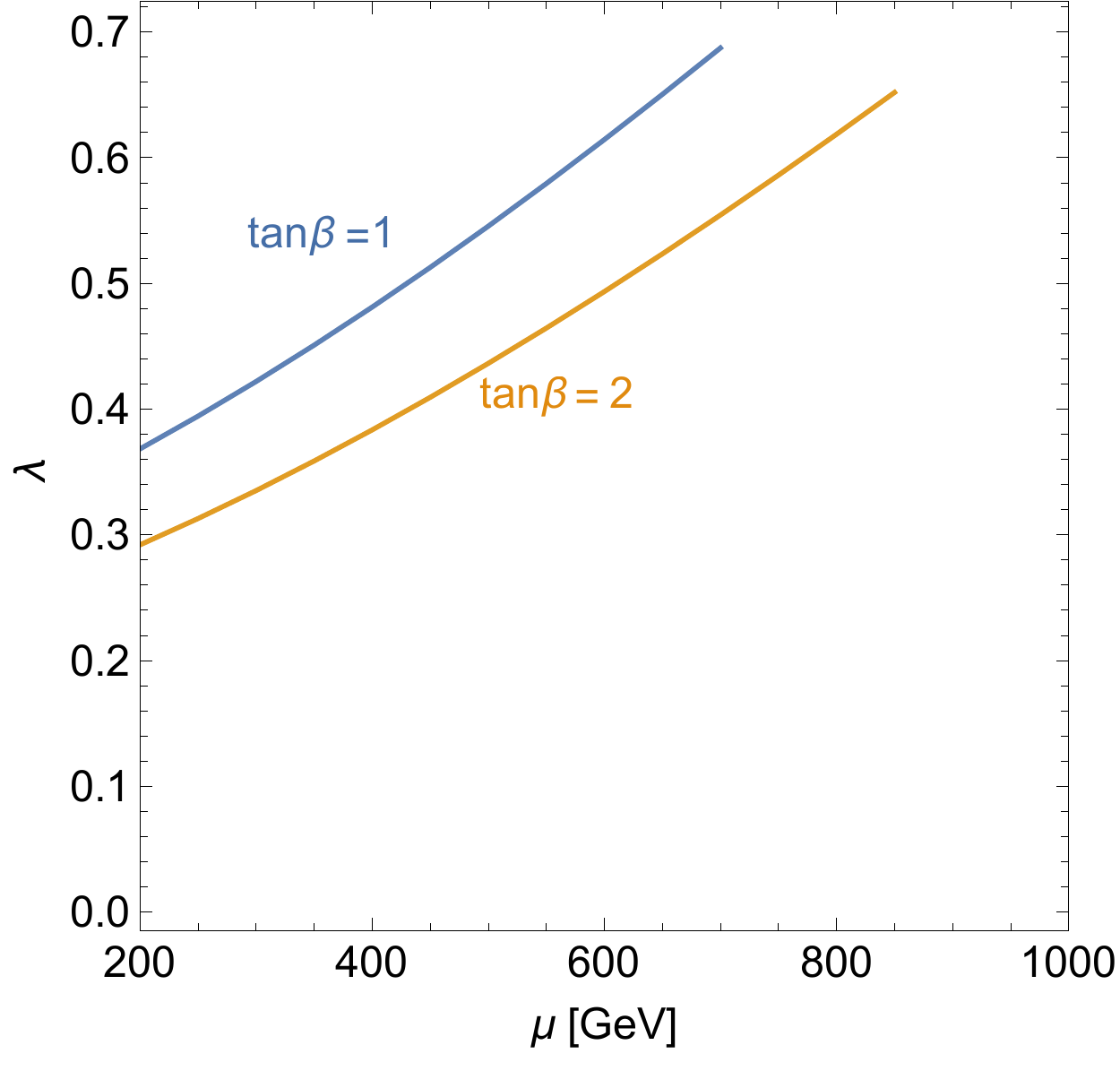}
\caption{Top row: Maximal values of $v_{\Delta}$ that allow $\lambda$ to set the Higgs mass to $125~\gev$ and yield a minimized potential as a function of $\mu$. Bottom: Value of $\lambda$ needed to attain the observed Higgs mass for $v_{\Delta}=10~\gev$. The triplet supersymmetric mass is set to $\mu_{\Sigma}=250~\gev$, and the other values are as in Eq.~\eqref{eqn:benchmark}.}
\label{fig:vdlambdaTB1max}
\end{center}
\end{figure}

By fixing the mass of the stops, the only way to alter the mass of the Higgs is through the remaining parameters, $\mu,\mu_{\Sigma},\lambda$, and $v_{\Delta}$. To study the effect of the triplet states on dark matter, we examine the case in which either the doublet- or the tripletlike fermions are lighter. We fix $\mu_{\Delta}=250~\gev$ and scan over the values of $\mu$. 

Recall that in order to achieve a minimum of the potential from the minimization conditions in Eqs.~(\ref{eqn:minh1})-(\ref{eqn:minSig0}), rather than a saddle point, there exist constraints on the relationship between $\mu$ and $\mu_{\Delta}$ beyond  Eq.~\eqref{eqn:custmincond}. However, the latter equation might give us some intuition on which $\mu$ and $\mu_{\Delta}$ values we can take since a saddle point at the origin forces the potential to have a minimum. When $\tan{\beta}=1$ and $A_\lambda=0$ the equation simplifies to $2\mu>\mu_\Delta$ and we see that we cannot look at regions where $\mu_{\Delta}$ is significantly heavier than $\mu$ and still minimize the potential.\footnote{Of course, this does not mean that triplets cannot be decoupled supersymmetrically. The limit $\mu_\Delta\to\infty$ yields the MSSM, in which case Eq.~\eqref{eqn:custmincond} does not apply.}

Once $\mu$ and $\mu_{\Delta}$ are fixed, we have to chose $\lambda$ and $v_{\Delta}$. We do this in two different ways: 
\begin{enumerate}
	\item To maximize the value of $v_{\Delta}$, we start with $v_{\Delta}=30~\gev$, which we take as the upper limit as suggested by the analysis of Ref.~\cite{Garcia-Pepin:2014yfa}. We then scan over $\lambda$ to set the Higgs mass (including radiative corrections). Once the lightest $CP$-even Higgs has a mass of $125~\gev$, we examine the rest of the spectrum. If other scalars have gone tachyonic, or the value of $\lambda$ needed is greater than 0.75, this value of $v_{\Delta}$ is excluded. We then lower $v_{\Delta}$ and repeat the process until a $125~\gev$ Higgs is obtained and the vacuum minimized. The resulting values of $v_{\Delta}$ and $\lambda$ are plotted in the top row of Fig.~\ref{fig:vdlambdaTB1max} over a range of $\mu$.
	\item The other option is to keep the value of $v_{\Delta}$ constant as we scan across $\mu$. The region of $\mu$ that can yield the correct Higgs mass and successfully minimize the potential is smaller for large values of $v_{\Delta}$. Because of this, we set $v_{\Delta}=10~\gev$ for our study of this method. The lower panel of Fig.~\ref{fig:vdlambdaTB1max} displays the values of $\lambda$ needed for both $\tan\beta=1$ and $\tan\beta=2$. Note that $\tan\beta=2$ needs smaller values of $\lambda$ because there are tree-level MSSM contributions to the Higgs mass. This allows for a larger range of $\mu$ than the $\tan\beta=1$ case.
\end{enumerate}

\begin{figure}
\begin{center}
\mbox{
\raisebox{-0.5\height}{\includegraphics[width=5.5in]{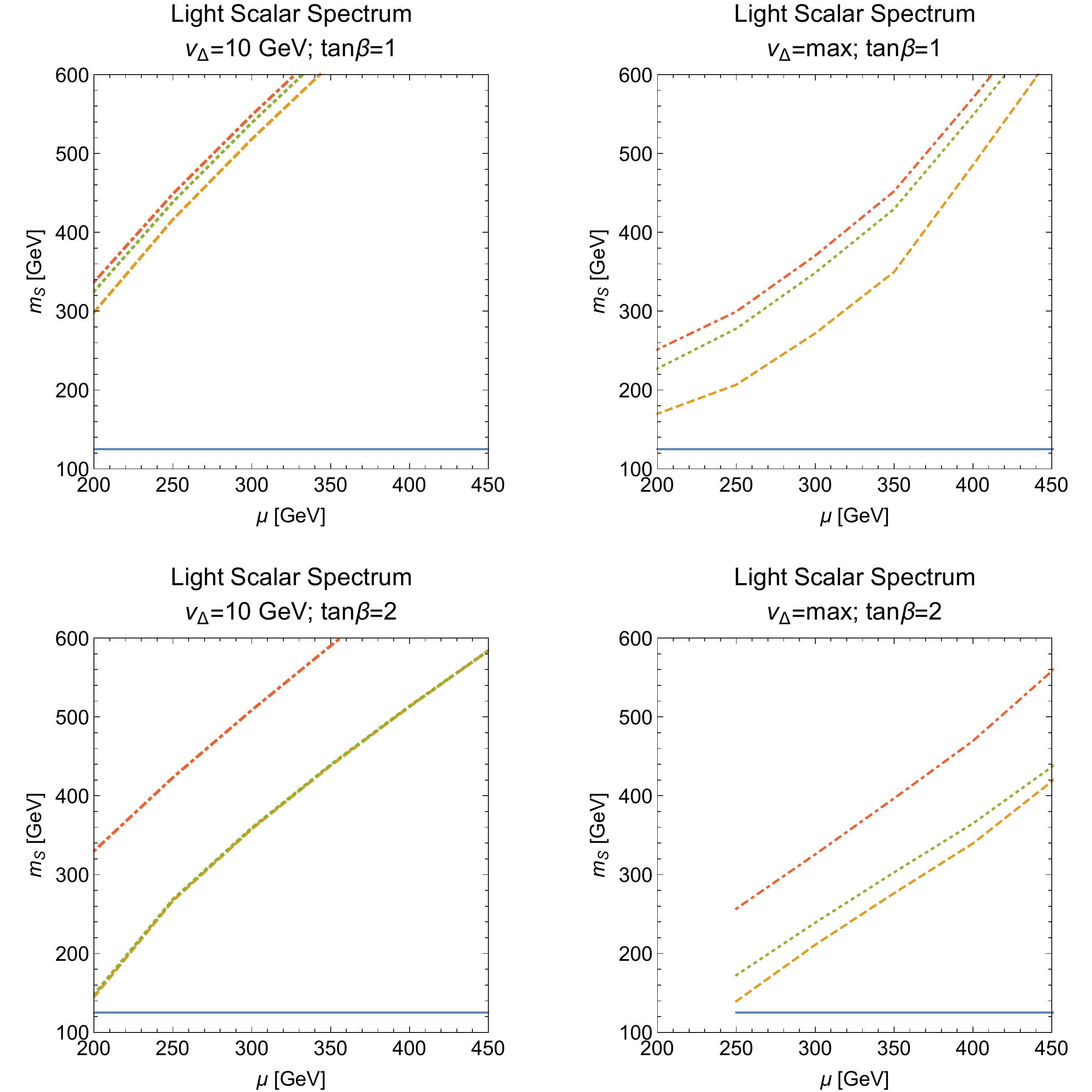}}
\raisebox{-0.5\height}{\includegraphics[scale=1]{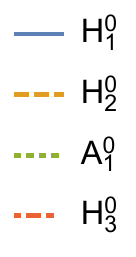}}
}
\caption{The left panels show the spectrum of the neutral light scalars when $v_{\Delta}=10~\gev$ and $\lambda$ is changed to set the Higgs mass.  The right panels use the maximum allowed value for $v_{\Delta}$ for each $\mu$ value. The upper (lower) panels contain $\tan\beta=1$ ($\tan\beta=2$). Changing $\tan\beta$ greatly affects the masses of $H_2^0$ and $A_2^0$, but $H_3^0$'s mass is similar for both choices.}
\label{fig:ConstantScalarSpec}
\end{center}
\end{figure}

The spectrum of the light neutral scalars is plotted in Fig.~\ref{fig:ConstantScalarSpec} for $v_{\Delta}=10~\gev$ and $v_{\Delta}$ maximized in the left and right panels respectively. When $\tan\beta=1$, shown in the upper panels, the scalars $H_2^0, A_1^0, \text{ and } H_3^0$ have similar masses, which increase as a function of $\mu$. The lightest that these scalars can be is $\sim300~\gev$. The other neutral scalars all have masses greater than $600~\gev$ and therefore are not shown in the plots. In the lower panels, the same spectra are shown for $\tan\beta=2$. In this case, both $H_2^0$ and $A_1^0$ are nearly degenerate in mass, and much lower in mass than when $\tan\beta=1$. This partially comes from the smaller value of $\lambda$ needed to raise the Higgs mass for $\tan\beta=2$. Conversely, the mass of $H_3^0$ does not change much between the two choices of $\tan\beta$. If the maximum $v_{\Delta}$ is chosen instead of using the constant $v_{\Delta}=10~\gev$, the masses of $H_2^0, A_1^0, \text{ and } H_3^0$ will drop. The separation of the states will also depend on $v_{\Delta}$ so increasing it helps to remove the degeneracy of the scalars. 

We do not perform any collider constraints on searches for these extra possible scalars. However, we see that the model allows for some to be very light. A dedicated search could therefore exclude large regions of parameter space in a quicker and more conclusive way than either Higgs precision measurements or direct detection experiments.

%*********************NeutralinoMixing***********************
\section{Neutralino Mixing}
\label{sec:NeutralinoMixing}
%*********************NeutralinoMixing***********************

The addition of three triplet chiral superfields adds to the neutralino content of the model three extra states. The mass Lagrangian in the basis $\psi^0 = \left( \tilde{B}^0 , \tilde{W}^3 , \tilde{H}_1^0 , \tilde{H}_2^0 , \tilde{\phi}^0 , \tilde{\chi}^0 , \tilde{\psi}^0 \right)$ is then
\begin{equation}
\Lc_{\text{neutralino mass}} = -\frac{1}{2} (\psi^0)^T \mathbf{M} \psi^0 + \text{c.c.},
\end{equation}
where
\begin{equation}
\mathbf{M} = 
\begin{pmatrix} 
M_1 & 0 & -\frac{1}{\sqrt{2}} g^{\prime}  c_{\beta} v_H & \frac{ \sqrt{2}}{2} g^{\prime} s_{\beta} v_H & 0 & - g^{\prime} v_{\Delta} & g^{\prime} v_{\Delta} \\
0 & M_2 & \frac{ \sqrt{2}}{2} g_2 c_{\beta} v_H & -\frac{1}{\sqrt{2}} g_2  s_{\beta} v_H & 0 & g_2 v_{\Delta} & - g_2 v_{\Delta} \\
-\frac{1}{\sqrt{2}} g^{\prime}  c_{\beta} v_H & \frac{1}{\sqrt{2}} g_2  c_{\beta} v_H & - \sqrt{2} \lambda v_{\Delta} & -\frac{1}{\sqrt{2}} \lambda v_{\Delta} - \mu & - \lambda s_{\beta} v_H & 0 & -2 \lambda c_{\beta} v_H \\
\frac{ \sqrt{2}}{2} g^{\prime} s_{\beta} v_H & -\frac{1}{\sqrt{2}} g_2  s_{\beta} v_H & -\frac{1}{\sqrt{2}} \lambda v_{\Delta} - \mu & -\sqrt{2} \lambda v_{\Delta} & - \lambda  c_{\beta} v_H & -2 \lambda s_{\beta} v_H & 0 \\
0 & 0 & - \lambda s_{\beta} v_H &  \lambda c_{\beta} v_H & \mu_{\Delta} & -\frac{1}{\sqrt{2}} \lambda_3 v_{\Delta} & -\frac{1}{\sqrt{2}} \lambda_3 v_{\Delta} \\
g^{\prime} v_{\Delta} & g_2 v_{\Delta} & 0 & -2 \lambda  s_{\beta} v_H & -\frac{1}{\sqrt{2}} \lambda_3 v_{\Delta} & 0 & \mu_{\Delta} - \frac{1}{\sqrt{2}} \lambda v_{\Delta} \\
g^{\prime} v_{\Delta} & -g_2 v_{\Delta} & -2\lambda  c_{\beta} v_H & 0 & -\frac{1}{\sqrt{2}} \lambda_3 v_{\Delta} & \mu_{\Delta} -\frac{1}{\sqrt{2}} \lambda_3 v_{\Delta} & 0
\end{pmatrix} 
\label{eqn:neutmass}
\end{equation}
and $s_{\beta}$ and $c_{\beta}$ are shorthand for the sine and cosine of $\beta$, respectively.

Overall, the masses are controlled by $M_1, M_2, \mu$, and $\mu_{\Delta}$ for the Bino, Wino, Higgsinos, and tripletinos, respectively. There are additional contributions to the masses and mixings scaling with either $v_H$ or $v_{\Delta}$. To provide a good dark matter candidate, we want the LSP to be the lightest neutralino; its composition will then determine the annihilation and direct detection cross sections.

\begin{figure}[t]
\begin{center}
 \raisebox{-0.5\height}{\includegraphics[width=5.5in]{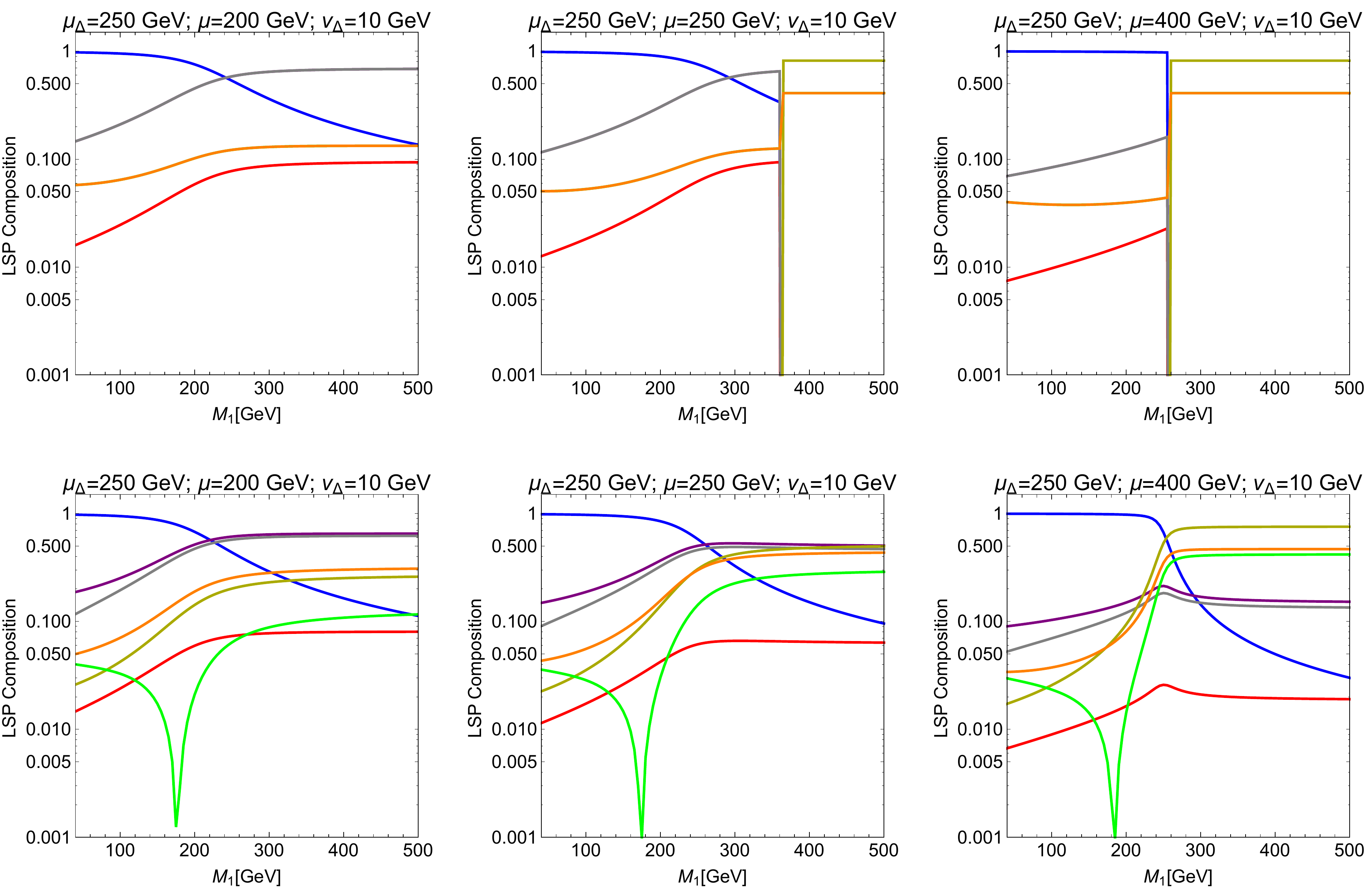}}
 \raisebox{-0.5\height}{\includegraphics[scale=1]{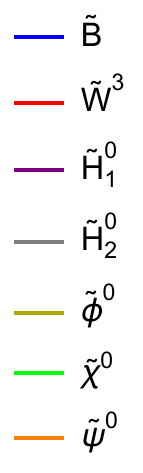}}
\caption{Composition of the LSP in terms of gauge eigenstates. The top row shows $\tan\beta=1$ and the bottom shows $\tan\beta=2$. The columns correspond to $\mu=(200,250,400)~\gev$, respectively, while the tripletino mass is set to $\mu_{\Delta}=250~\gev$. The Wino has been decoupled with $M_2=1~\tev$.  Note in the top middle and top right plots the presence of a tripletlike eigenvalue, which is totally decoupled from the rest of the neutralino mass matrix, made out of only $\tilde{\psi},\tilde{\phi}$, and $\tilde{\chi}$. It corresponds to an $SU(2)_V$ 5-plet in the custodial basis.
}
\label{fig:comp}
\end{center}
\end{figure}

The composition of the LSP in terms of the gauge eigenstates is shown in Fig.~\ref{fig:comp} for the case in which the VEV of the triplets is constant $10~\gev$ and $\tan\beta=1(2)$ in the top (bottom) row. The left panels have the Higgsino-like states lighter than the tripletino ones, using $\mu=200~\gev$ and $\mu_{\Delta}=250~\gev$. The middle panel has both the Higgsino and tripletino masses set to $\mu=\mu_{\Delta}=250~\gev$. Finally, the right panel examines when the triplet states are lighter than the Higgsino, with $\mu=400~\gev$ and $\mu_{\Delta}$ still at $250~\gev$. To simplify the situation as much as possible, we decouple the Wino by setting $M_2=1~\tev$.

In the custodial situation, the doublet components of the LSP are equal and the triplet components are separately equal over most of the parameter space. The $\tan\beta=2$ case has each Higgsino and tripletino contributing differently to the LSP. Despite the complexity of the plots, there are a few overarching trends. 

In Sec.~\ref{sec:NeutralinoDM}, we argued that the Bino component of the LSP must dominate in order to achieve the correct relic abundance of dark matter. The interesting regions to examine in the compositions plots are then $M_1 < \mu,\mu_{\Delta}$. In this region, even when $\mu>\mu_{\Delta}$, the second-largest component of the LSP is Higgsino rather than tripletino which is true even for quite large values of the Higgsino mass. This is due to the mixing of the Bino with the Higgsinos or tripletinos, which comes from off-diagonal terms weighted with $v_H$ or $v_{\Delta}$, respectively. Because of Eq.~(\ref{vev246}), $v_{H}\gg v_{\Delta}$, and the Higgsino mass needs to be much larger than the tripletino mass in order for the triplet contribution to the LSP to be larger than the Higgsino component. So even though the mass of the Higgsino can be larger than the tripletino mass, the mixing of the Bino with the Higgsino can be what causes the correct annihilation rate.

As $\mu$ is further increased, the amount of Higgsino in the LSP drops past the point where mixing alone can yield the correct relic abundance. Looking only at regions where $M_1 < \mu_{\Delta}$, we see that the triplet states do not contribute much to the LSP. By removing the Higgsino, the LSP is made more pure Bino, rather than increasing the triplet amount. The only possibility of well tempering for this will then require coannihilations of the Bino-like LSP with a tripletlike state.

%*********************darkmatter***********************	
\section{dark matter}
\label{sec:darkmatter}
%*********************darkmatter***********************	

To examine the dark matter of the SCTM the model was implemented into \textsc{SARAH} \cite{Staub:2008uz, Staub:2009bi, Staub:2011dp, Staub:2013tta, Staub:2015kfa}. With this, a code was generated for \textsc{SPheno} \cite{Porod:2003um, Porod:2011nf} and \textsc{CalcHep} \cite{Pukhov:1999gg}. The \textsc{SPheno} code calculates the spectrum, outputting a parameter card that can be read by \textsc{MicrOMEGAs 3} \cite{Belanger:2013oya}. The program \textsc{MicrOMEGAs 3} uses the \textsc{CalcHep} code to calculate the dark matter properties. 

\subsection{Thermal relic density}

For each of the choices of $\tan\beta$ and the method of picking $v_{\Delta}$, we scan over the possible $\mu$ values for $\mu_{\Delta}=250~\gev$, using $50~\gev$ step sizes. At each point in $\mu$, we then scan over $M_1$ to find the Bino masses that yield the correct relic abundance of dark matter. We start with $M_1 = 40~\gev$ and take $1~\gev$ steps until $M_1 > 100~\gev$, at which point a $5~\gev$ step is used to save on computing time.

\begin{figure}[t]
\begin{center}
\includegraphics[width=0.4\linewidth]{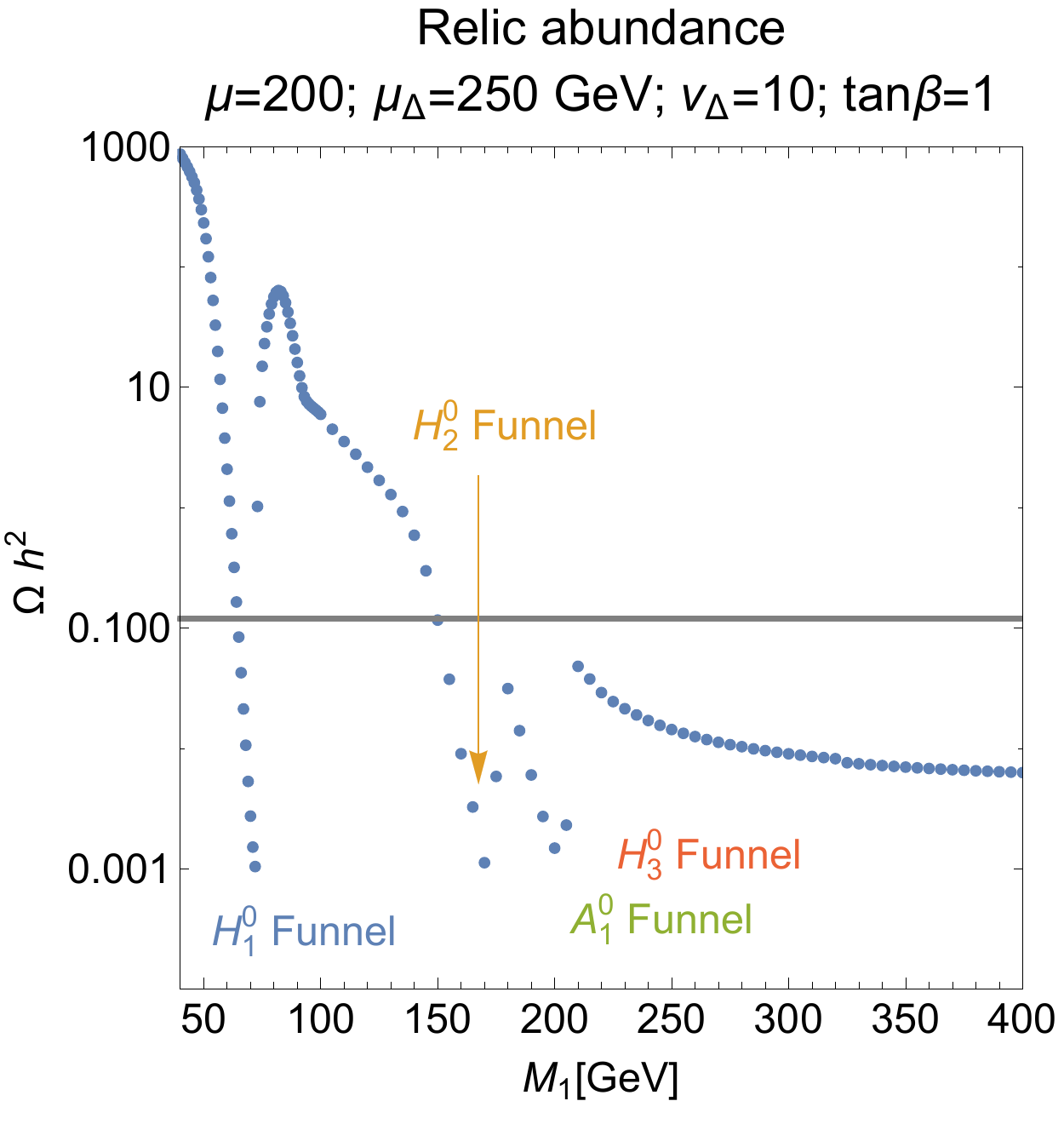}
\caption{Relic abundance for the model with $\mu=200~\gev$, $\mu_{\Delta}=250~\gev$, $v_{\Delta}=10~\gev$, and $\tan\beta=1$. The gray line marks the observed relic abundance in the Universe today. As the mass of the LSP crosses over half the mass of one of the scalars in the model, the annihilation cross section greatly increases, leading to lower relic abundances. When the LSP mass gets close to the mass of the Higgsino, the mixing and coannihilations take over and the relic abundance stays below the observed value.}
\label{fig:dmExample}
\end{center}
\end{figure}

Figure~\ref{fig:dmExample} shows an example of the relic abundance calculated at each $M_1$ value for the point $\mu=200~\gev$, $v_{\Delta}=10$, and $\tan\beta=1$. The gray line marks $\Omega h^2 = 0.1187$, the observed relic abundance in the Universe~\cite{Agashe:2014kda}. The scalar masses do not depend on the $M_1$ value and are given by
\be
\begin{aligned}
m_{H_1^0} &= 125~\gev, ~~ m_{H_2^0} = 299~\gev, \\
m_{A_1^0} &= 325~\gev, ~~m_{H_3^0} = 337~\gev, ~~ \text{and others}>700~\gev.
\end{aligned}
\ee
Three dips in the relic abundance are seen in the plot corresponding to the $H_1^0$ funnel, the $H_2^0$ funnel, and one for the nearly degenerate $A_1^0$ and $H_3^0$ states occurring when the Bino mass is roughly half the scalar mass. There are three $M_1$ values of this model point that yield the correct relic abundance. The first two correspond to going into and out of the lightest Higgs funnel, and the third one is at the start of the $H^0_2$ funnel. However, the next funnels corresponding to $A_1^0$ and $H_3^0$ are close together, so the effect of having multiple nearly resonant $s$-channel annihilations keeps the relic abundance below the observed value. This runs into the region where $M_1 > \mu$ and the Higgsino becomes the LSP, leaving not enough dark matter in the current Universe.

For each $\mu$ value in our model scans, we do the same process. Whenever the relic abundance at one $M_1$ value crosses from one side of the observed value to the other at the next $M_1$ step, we do a more dedicated scan to find the $M_1$ value to a higher degree of accuracy. We then classify the point according to the process that is driving the annihilations by comparing the LSP mass to half the mass of the scalars or $10\%$ higher than the LSP mass with that of the next-lightest electroweakino. The piece giving the minimum of
\be
\text{min}\left( \left|m_{\tilde{\chi}^0_1} - \frac{m_{H^0_1}}{2}  \right| ,  \left|m_{\tilde{\chi}^0_1} - \frac{ m_{H^0_2}}{2} \right| , \left|m_{\tilde{\chi}^0_1} - \frac{ m_{A^0_1}}{2} \right| ,  \left|m_{\tilde{\chi}^0_1} - \frac{ m_{H^0_3}}{2} \right|, 	\left| m_{\tilde{\chi}^{0(\pm)}_{\text{NLSP}}} -1.1 \times m_{\tilde{\chi}^0_1} \right|  \right)
\ee
yields a classification of the given scalar funnel or well tempering. This classification is only an approximation of what is actually causing the annihilations. In the nonrelativistic limit, annihilations through scalars occur through the $p$-wave, while pseudoscalars occur through the $s$-wave. Thus, when $A_1^0$ is close in mass to either $H_2^0$ or $H_3^0$, the classification scheme could point to the scalar instead of the pseudoscalar, even though the pseudoscalar contribution is larger. In addition, when the funnels are close to the well tempered region both process can be responsible for the annihilation. 

\begin{figure}[t]
\centering
\mbox{
\raisebox{-0.5\height}{\includegraphics[width=5.5in]{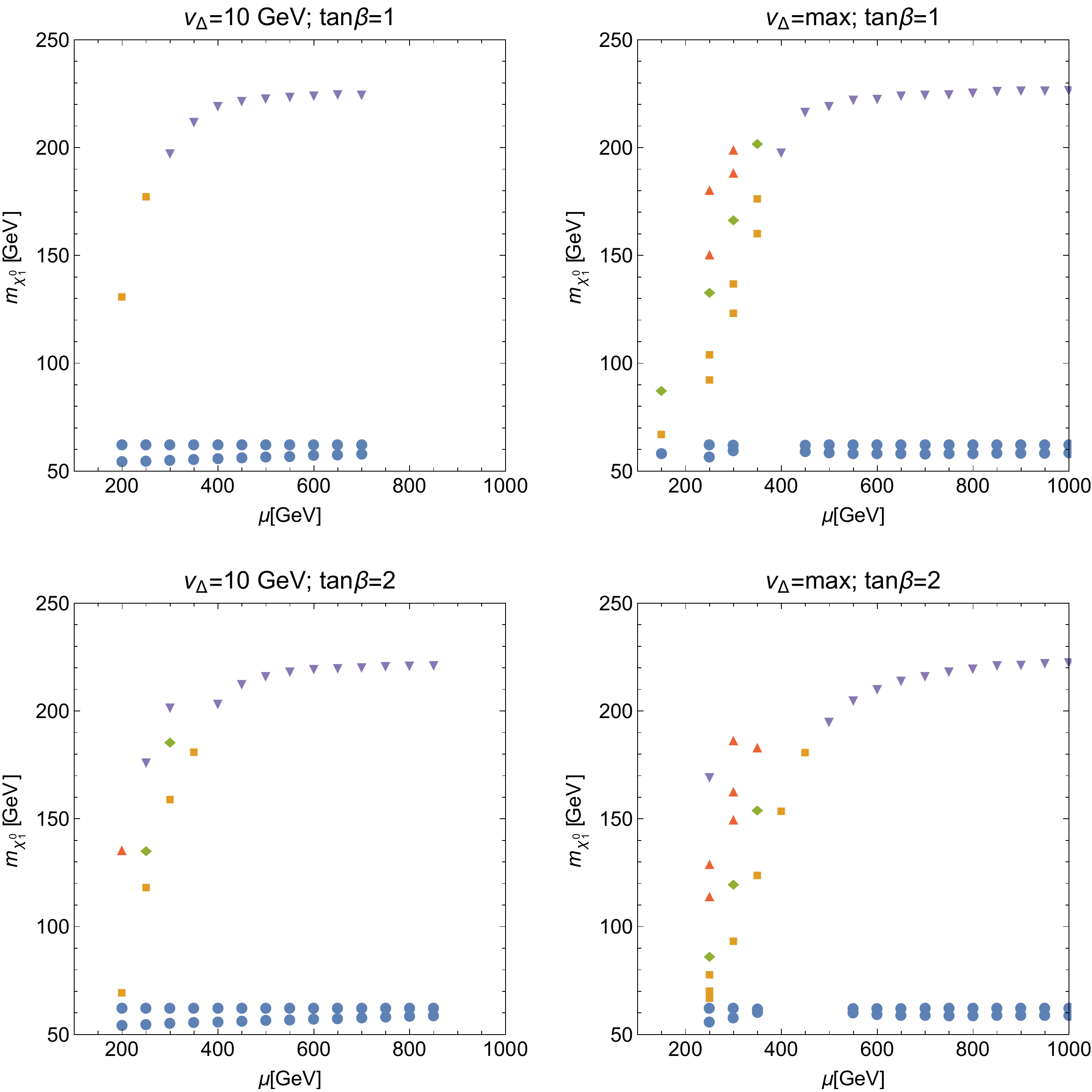}}
\raisebox{-0.5\height}{\includegraphics[scale=1]{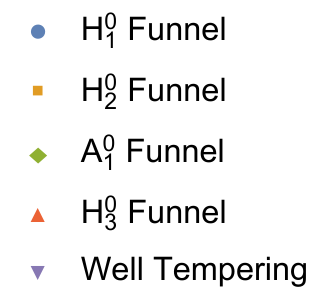}}
}
\caption{Points that yield the correct relic abundance of dark matter. The upper row is for the custodial case, while the lower has $\tan\beta=2$. The left panels keep $v_{\Delta}$ constant, and the right panels use the maximum allowed value for $v_{\Delta}$ for each $\mu$ value. The points are labelled corresponding to which annihilation channel dominates in the early Universe.}
\label{fig:dmM1Mu}
\end{figure}

The results of the classifications are plotted in Fig.~\ref{fig:dmM1Mu} for the different model choices in the $m_{\tilde{\chi}^0_1}$ vs. $\mu$ plane. The LSP is mostly Bino, so $M_1 \sim m_{\tilde{\chi}^0_1}$. The triplet scalars can be very light for $\tan\beta=2$ or if $\tan\beta=1$ when the VEV of the triplets takes on the maximum value allowed. Recall that Fig.~\ref{fig:ConstantScalarSpec} shows that these masses increase as a function of $\mu$. As such, the funnels for the Tripletlike, $H^0_2$, $A^0_1$, and $H^0_3$, scalars smoothly transition up to the point where well tempering happens at a lighter mass than needed for a triplet funnel. 

For every model choice examined, there is an $M_1$ value that will yield the correct relic abundance either through a Tripletlike scalar or well tempering. When $\mu$ is large enough that the triplet scalars funnels are not possible, the Higgsinos are heavy enough that the well tempering is not caused by Bino-Higgsino mixing but instead by coannihilations with the triplet fermions. \textit{Thus each model point examined is capable of setting the correct relic abundance using particles beyond the MSSM content.}

The large VEV of the triplets allows for the triplet scalars to be light. The lightness of these scalars is what allows the model points examined to always be able to set the relic abundance using either the triplet scalar funnels or the triplet fermions. However, lowering the triplet VEV, $v_{\Delta}$, raising the triplet supersymmetric mass, $\mu_{\Delta}$, or lowering the Wino mass, $M_2$, can disturb the possibility of achieving the correct relic abundance through a triplet state. The MSSM limit of the model takes the VEV of the triplets to zero. In this case, the triplet scalar soft masses go to infinity and do not contribute to the annihilations.\footnote{This also happens in triplet models in which the $\rho$ parameter is not protected by a custodial symmetry, as the triplet extension of the MSSM~\cite{Arina:2014xya}, and $v_\Delta$ is strongly constrained by electroweak precision observables.} The Higgsino alone satisfies the correct relic abundance if its mass is $\sim1.1~\tev$. As such, if $\mu_{\Delta}$ is much larger than that, the triplet fermions cannot play a role in well tempering. Such a large value of $\mu$ would also keep the triplet scalars heavy, so such a case would have no way of using the triplet superfield to set the relic abundance. Finally, the Wino has been raised above the mass of the Higgsinos and tripletinos for this study. Bino-Wino well tempering can also be done if $M_1 \simeq M_2 < \mu,\mu_{\Delta}$. In this case the relic abundance could be set before the triplets have a chance to affect things.

\subsection{Direct detection}

There have been many experimental searches for the direct detection of dark matter. For the mass ranges considered here, the Particle Data Group~\cite{Agashe:2014kda} shows that the best limits are currently coming from the LUX Collaboration~\cite{Akerib:2013tjd} for spin-independent searches and the COUPP Collaboration~\cite{Behnke:2008zza, Behnke:2010xt} for spin-dependent measurements. Super-Kamiokande~\cite{Tanaka:2011uf} and IceCube~\cite{IceCube:2011aj, Aartsen:2012kia} have better spin-dependent exclusions, but are indirect constraints that rely on the annihilation of dark matter in the current Universe and depend on the the byproducts of the annihilation that change as the LSP composition changes. We then only compare our results with the LUX and COUPP constraints.

\begin{figure}[t]
\begin{center}
\mbox{
\raisebox{-0.5\height}{\includegraphics[width=5.5in]{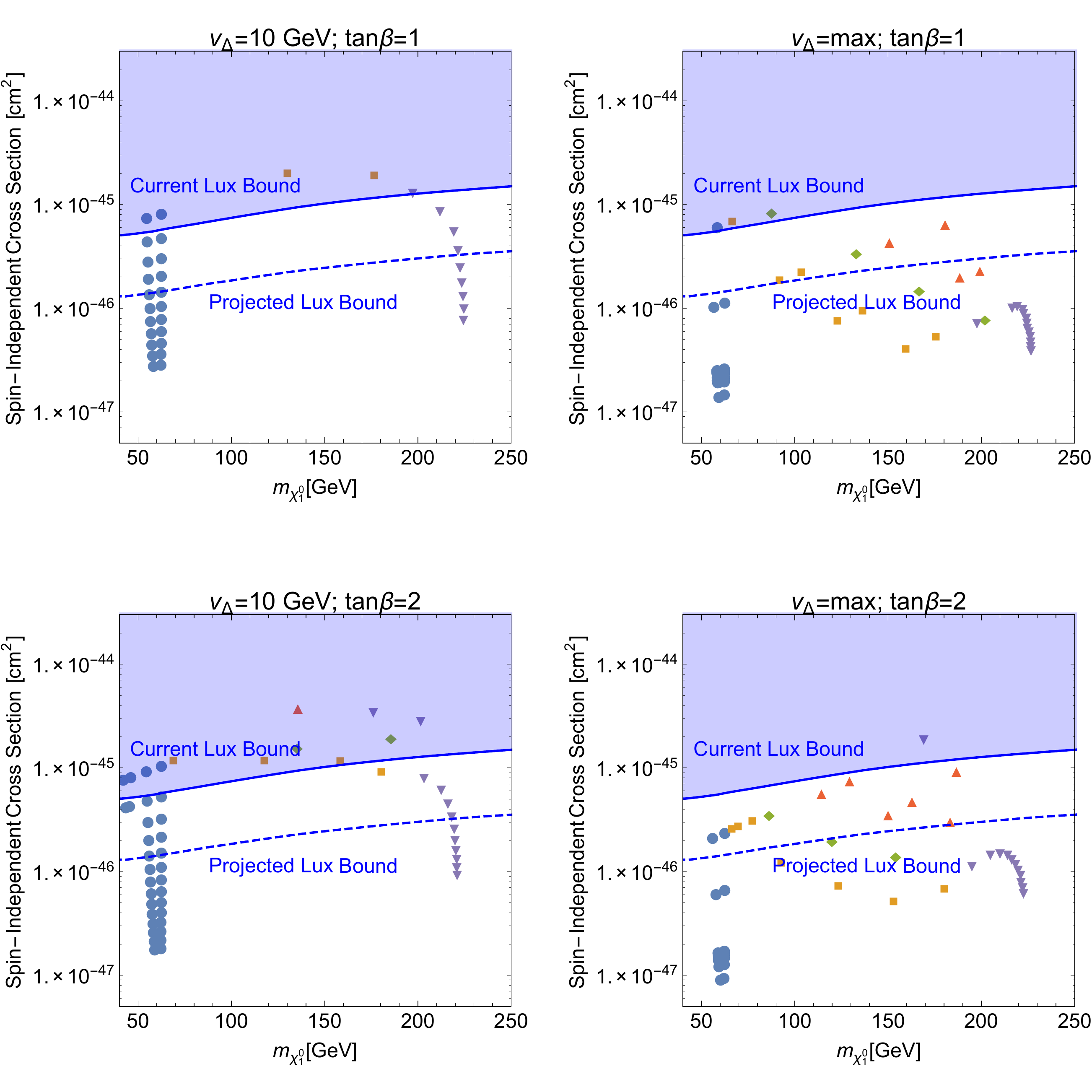}}
\raisebox{-0.5\height}{\includegraphics[scale=1]{TypeLegend}}
}
\caption{Spin-independent dark matter nucleon cross sections. Each point meets the correct relic abundance with the annihilation mode marked. Points with smaller Higgsino components have a lower spin-independent cross section.}
\label{fig:dmSI}
\end{center}
\end{figure}

The spin-independent cross sections for the points satisfying the correct relic abundance are shown in Fig.~\ref{fig:dmSI}. The \textsc{micrOMEGAs 3} output provides both the cross section of the dark matter with a proton and a neutron; we take the maximum of these. The points are marked in the same fashion as Fig.~\ref{fig:dmM1Mu} to show how the relic abundance is being achieved. The upper (lower) panels show $\tan\beta=1\, (2)$ while the left and right panels display $v_{\Delta}=10~\gev$ and when $v_{\Delta}$ is maximized at each point, respectively. The shaded blue region is excluded by the LUX bound, and the dashed blue line is the projected sensitivity of LUX.

The spin-independent cross section is mediated by the doublet scalars. There is not much difference between the $\tan\beta=1$ and $\tan\beta=2$ models in terms of the cross sections. For $v_{\Delta}=10~\gev$, both have a region where the dark matter mass is between 100 and 200 GeV which can be excluded by LUX. The points are achieved through a triplet funnel, and to get masses in this range for the LSP, the values of $\mu$ are low. Referring back to Fig.~\ref{fig:comp}, low values for $\mu$ and $M_1$ give the LSP a moderate Higgsino component. This Higgsino component is what drives the nuclear cross sections to be so large. The cross sections are lower when the maximum value of $v_{\Delta}$ is used. In this case, there are few points that are currently excluded by LUX. The larger value of $v_{\Delta}$ lowers the masses of the Tripletlike scalars. This pushes the triplet funnels and the well-tempering regions to larger values of $\mu$, further decreasing the Higgsino component and the spin-independent cross section. Fortunately, there are still many points that can be probed by LUX in the future. However, the points which are well tempered through Bino-tripletino coannihilations remain under the projected bound, due to the minimal Higgsino component of the LSP. 

\begin{figure}[t]
\begin{center}
\mbox{
\raisebox{-0.5\height}{\includegraphics[width=5.5in]{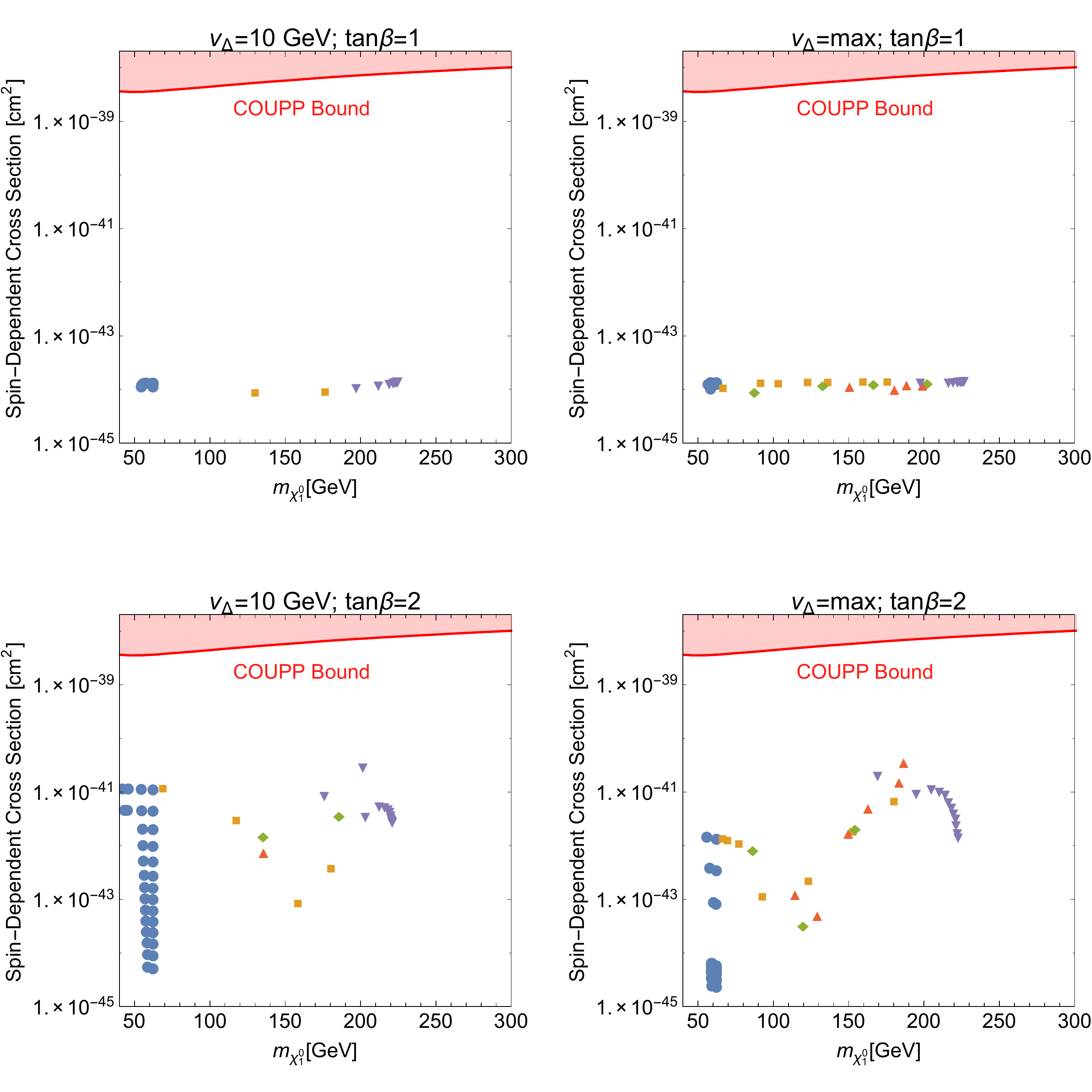}}
\raisebox{-0.5\height}{\includegraphics[scale=1]{TypeLegend}}
}
\caption{Spin-dependent dark matter nucleon cross sections. Each point meets the correct relic abundance with the annihilation mode marked. The parity-violating $Z$ couplings vanish in the custodial case.}
\label{fig:dmSD}
\end{center}
\end{figure}

The spin-dependent interactions are mediated by the $Z$ boson and the cross sections are shown in Fig.~\ref{fig:dmSD}. The panels use the same labelling as Figs.~\ref{fig:dmM1Mu} and \ref{fig:dmSI}. In the custodial case, with $\tan\beta=1$, the mass eigenstates of both the fermions and the scalars of the Higgs doublet and triplet superfields form representations of  $SU(2)_V$. The parity-violating $Z$ coupling therefore vanishes in this case. And while this is also true in the MSSM for $\tan\beta=1$, the SCTM provides motivation for this choice of $\tan\beta$. The model points examined for $\tan\beta=2$ no longer have vanishing $Z$ couplings with the LSP. The cross sections are much larger in this case, particularly for the well-tempered points, which have low spin-independent cross sections. However, even these large cross sections are still $\sim 2$ orders of magnitude below the COUPP bound.

\subsection{Indirect detection}

The direct detection experiments rely on dark matter interacting with detectors on Earth. It is also possible to observe astrophysical objects containing large dark matter densities. In these regions of space, the LSP can still annihilate. The annihilation does not occur through a diphoton process, which would lead to a monochromatic signal. Instead, experiments must search for photons coming from the byproducts of the annihilation.

The annihilation cross section in the current Universe can be much different than in the early Universe. Scalar funnels (not pseudo) are velocity suppressed in the nonrelativistic limit. As the temperature has cooled since freeze-out, the annihilations proceeding through scalars should be significantly smaller than the $\sim3\times10^{-26} \text{cm}^3/\text{sec}$ needed at freeze-out. For the well tempering through coannihilations, the coannihilating particle is no longer around in the current Universe, so we expect the annihilation cross section to be lower now as well. 

\begin{figure}[t]
\begin{center}
\mbox{
\raisebox{-0.5\height}{\includegraphics[width=5.5in]{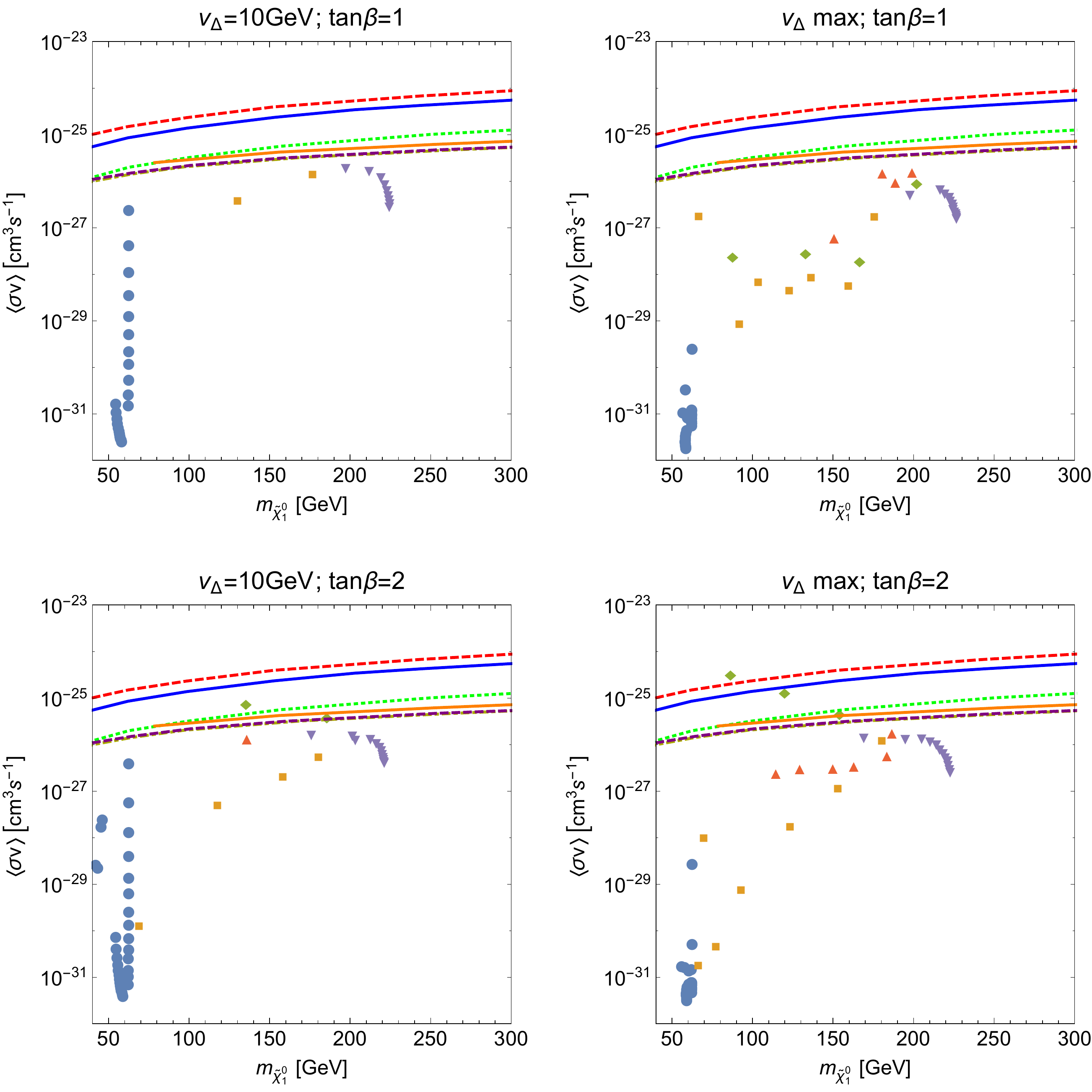}}
\raisebox{-0.5\height}{\includegraphics[scale=1]{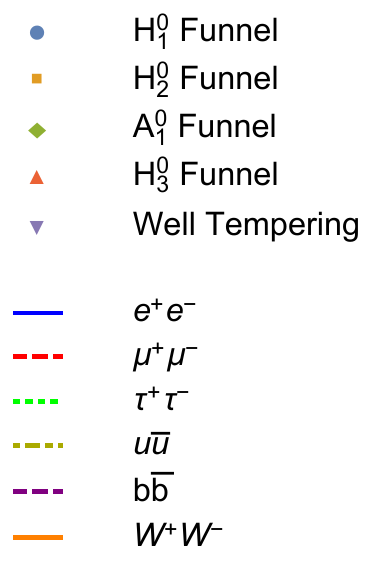}}
}
\caption{Annihilation cross section times velocity of dark matter in the Galaxy in the current Universe. Each point meets the correct relic abundance with the annihilation mode in the early Universe marked. The lines mark the limits assuming the annihilation occurs $100\%$ of the time through the given channel, each resulting in different spectra of photons measured here on Earth.}
\label{fig:dmIndirectD}
\end{center}
\end{figure}

The Fermi-LAT Collaboration \cite{Ackermann:2013yva,Ackermann:2015zua} has placed limits on the annihilation cross section of dark matter from the observation of satellite galaxies. The limits are framed in the context of the annihilations proceeding $100\%$ of the time through either the $e^+e^-$, $\mu^+\mu^-$, $\tau^+ \tau^-$, $u\bar{u}$, $b\bar{b}$, or $W^+W^-$ channel. In Fig.~\ref{fig:dmIndirectD}, the resulting limits are plotted with our model points yielding the correct relic abundance. The upper (lower) panels show $\tan\beta=1\ (2)$, while the left and right panels display $v_{\Delta}=10~\gev$ and when $v_{\Delta}$ is maximized at each point, respectively.  \textcolor{red}{A few} points for the $\tan\beta=2$ case are possibly excluded by these searches. However, these each have the largest annihilation channel being $\tilde{\chi}^0_1 \tilde{\chi}^0_1 \rightarrow H_1^0 Z$. The spectrum of photons coming from the decays of the $H^0_1$ and $Z$ will not map directly onto any of the Fermi-LAT limits. The fact that the SCTM has more neutral Higgs funnels opens this possibility of having different annihilation modes. A more detailed study would therefore be needed in order to conclusively exclude points from the SCTM due to indirect constraints.

We also note that some of the points marked as annihilating through the pseudoscalar $A_1^0$ have particularly large annihilations in the current Universe. These interesting points have $A_1^0$ very close in mass to either $H^0_{2}$ or $H^0_{3}$, and there are interference effects in the early universe keeping the annihilation cross section small enough. In the current Universe, when the scalars do not play as much of a role, the annihilations proceed with less interference. Similarly, many points marked as $H^0_3$ funnels seem to have annihilation rates larger than expected in the current Universe. If the rates are scaled up by the larger velocity at freeze-out, the annihilation rate would seem to be too large. However, these points lie close to the well-tempered region, so it is likely that a simple classification does not work well for points where both processes are important.

%*********************Discussion***********************	
\section{Discussion and Conclusions}
\label{sec:Discussion}
%*********************Discussion***********************

We have studied a supersymmetric model in which the Higgs sector of the superpotential is extended by three $SU(2)_L$ triplet fields and is manifestly invariant under $SU(2)_L \otimes SU(2)_R$. When the triplet VEVs are aligned, the custodial symmetry of this setup allows the triplet fields to be light and develop large vacuum expectation values with no effect on the $\rho$ parameter but participating in the electroweak symmetry breaking. Therefore, we only allow this custodial setup to be broken by $\tan\beta$, the ratio of the doublet VEVs. This has an effect on the dark matter phenomenology, but does not alter the electroweak precision constraints. The $F$ terms coming from the triplet fields help to raise the mass of the Higgs to its observed value, lowering the needed mass of the stops.

We studied the case in which the lightest supersymmetric particle is a neutralino, which with R parity gives a stable dark matter candidate. For the dark matter candidate to yield the correct relic abundance, it must have a large Bino component to not annihilate too quickly in the early Universe. Well tempering mixes the Bino with either the Higgsino, Wino, or tripletino in just the right amount to give the observed relic abundance. If the Bino component is too large, dark matter does not annihilate quickly enough in the early Universe, unless the mass of the dark matter particle is about half the mass of a boson.  We found that the triplet scalars or the triplet fermions can play a role in the annihilation of dark matter in the early Universe over a large range of values for the Higgsino mass parameter $\mu$. We compared the model points giving the correct relic abundance with the current best direct detection limits. The points with low $\mu$ values have at least a moderate Higgsino component and have either been excluded already or can be discovered in future results. At large values of $\mu$, the light triplet states still provide an efficient means of annihilating the dark matter, but hope of a direct detection of dark matter is lost. This is motivation for a detailed study of the LHC phenomenology of these models. 

For the study of dark matter, we were only concerned with the neutral triplet states. However, the triplets contain charged states. In fact, the second-lightest {\it CP}-even Higgs, $H^0_2$, is close in mass to both the lightest charged and doubly charged scalars. In addition, the mixing of the Tripletlike fermions leaves charged and doubly charged states very near in mass to the lightest neutral one. If the relic abundance of dark matter relies on these light states, they should be accessible at the LHC. While a detailed study is beyond the scope of this paper, a dedicated study of methods for searching for doubly charged fermions and scalars could offer valuable constraints on models such as these with exotic particle content.

To conclude, the SCTM helps raise the mass of the Higgs through extra $F$ terms. There are large regions where the lightest Higgs is consistent with current observations, and able to explain any deviations in future measurements. It also offers new methods to annihilate dark matter in the early Universe through triplet fermion coannihilations or triplet scalar portals. The triplet fermions have weak detector bounds, if any, as compared to coannihilations with squarks or sleptons. The triplet scalar funnels can also be light, and have charged partners making them easier to search for than plain MSSM neutral Higgs funnels. Only a dedicated LHC search for the triplet fermions or scalars could offer the most conclusive constraints on the model.

%*********************acknowledgments***********************	
\begin{acknowledgments}
The authors would like to thank Florian Staub for assistance with~\textsc{SARAH}. MGP would also like to thank Victor Martin-Lozano for his constant help during the course of this work. This research was supported in part by the Notre Dame Center for Research Computing through computing resources, by the National Science Foundation under Grant No. PHY-1215979, by the Spanish Consolider-Ingenio 2010 Programme CPAN
(CSD2007-00042), by CICYT-FEDER-FPA2011-25948, by the Severo Ochoa
excellence program of MINECO under the grant SO-2012-0234 and by
Secretaria d'Universitats i Recerca del Departament d'Economia i
Coneixement de la Generalitat de Catalunya under Grant 2014 SGR 1450. 
The work of MGP and MQ was partly done at IFT and ICTP-SAIFR (Sao Paulo, Brazil) under CNPq grant 405559/2013-5, and at the CERN Theory Division (Geneva, Switzerland).
\end{acknowledgments}
%*********************acknowledgments***********************	

%*********************Appendix***********************	
\appendix
%*********************Appendix***********************	

%*********************MinCon***********************	
\section{Minimization conditions}
\label{sec:MinCon}
%*********************MinCon***********************

From $\frac{\partial V}{\partial h1}|_{\frac{v_1}{\sqrt{2}}}=\frac{\partial V}{\partial h2}|_{\frac{v_2}{\sqrt{2}}}=\frac{\partial V}{\partial\phi^0}|_{\frac{v_\phi}{\sqrt{2}}}=\frac{\partial V}{\partial\psi}|_{\frac{v_\psi}{\sqrt{2}}}=\frac{\partial V}{\partial\chi}|_{\frac{v_\chi}{\sqrt{2}}}=0$ we get,
\be
\begin{aligned}
&m_1^2\equiv m_{H_1}^2+\mu^2 = \sqrt{2}v_\Delta(\lambda\mu-A-\lambda\mu_\Delta)-\frac{v_\Delta^2}{2}(2\lambda_3\lambda+5\lambda^2)
\\ &+ t_\beta\{m_3^2+\frac{v_\Delta}{\sqrt{2}}(4\lambda\mu-A-\lambda\mu_\Delta)-\frac{v_\Delta^2}{2}(\lambda_3\lambda+4\lambda^2)\}  
-\frac{g_1^2+g_2^2}{4}v_H^2c_{2\beta}-v_H^2\lambda^2(c_\beta^2+1)\label{eqn:minh1}  
\end{aligned}
\ee
\be
\begin{aligned}
&m_2^2\equiv m_{H_2}^2+\mu^2 = \sqrt{2}v_\Delta(\lambda\mu-A-\lambda\mu_\Delta)-\frac{v_\Delta^2}{2}\left(2\lambda_3\lambda+5\lambda^2\right)  \\ & +\frac{1}{t_\beta}\{m_3^2+\frac{v_\Delta}{\sqrt{2}}(4\lambda\mu-A-\lambda\mu_\Delta)-\frac{v_\Delta^2}{2}(\lambda_3\lambda+4\lambda^2)\} -\frac{g_1^2+g_2^2}{4}v_H^2c_{2\beta}-v_H^2\lambda^2(s_\beta^2+1) \label{eqn:minh2} 
\end{aligned}
\ee
\be
\begin{aligned}
&m_{\Sigma_1}^2=-\{B_\Delta+\mu_\Delta^2+\frac{v_\Delta}{\sqrt{2}}\left(A_3+3\lambda_3\mu_\Delta\right)+\frac{1}{\sqrt{2}}\frac{v_H^2}{v_\Delta}\left(2Ac_\beta^2+2\lambda\mu_\Delta s_\beta^2-4c_\beta s_\beta\lambda\mu\right)\\&+\frac{v_H^2}{2}\left(-(g_1^2+g_2^2)c_{2\beta}+2c_\beta s_\beta(\lambda_3\lambda+2\lambda^2)+2s_\beta^2\lambda_3\lambda+8c_\beta^2\lambda^2\right)+v_\Delta^2\lambda_3^2\} \label{eqn:minSig1}
\end{aligned}
\ee
\be
\begin{aligned}
&m_{\Sigma_{-1}}^2=-\{B_\Delta+\mu_\Delta^2+\frac{v_\Delta}{\sqrt{2}}\left(A_3+3\lambda_3\mu_\Delta\right)+\frac{1}{\sqrt{2}}\frac{v_H^2}{v_\Delta}\left(2As_\beta^2+2\lambda\mu_\Delta c_\beta^2-4c_\beta s_\beta\lambda\mu\right)\\&+\frac{v_H^2}{2}\left(-(g_1^2+g_2^2)c_{2\beta}+2c_\beta s_\beta(\lambda_3\lambda+2\lambda^2)+2c_\beta^2\lambda_3\lambda+8s_\beta^2\lambda^2\right)+v_\Delta^2\lambda_3^2\} \label{eqn:minSigm1}
\end{aligned}
\ee
\be
\begin{aligned}
&m_{\Sigma_0}^2=-\{B_\Delta+\mu_\Delta^2+\frac{v_\Delta}{\sqrt{2}}\left(A_3+3\lambda_3\mu_\Delta\right)
\\&+\frac{1}{\sqrt{2}}\frac{v_H^2}{v_\Delta}\left(2c_\beta s_\beta(A+\lambda\mu_\Delta)-2\lambda\mu\right)+\frac{v_H^2}{2}\left(c_\beta s_\beta 8 \lambda^2+(2\lambda_3\lambda+2\lambda^2)\right)+v_\Delta^2\lambda_3^2\} \label{eqn:minSig0}
\end{aligned}
\ee
and by making $\tan{\beta}\rightarrow 1$ we recover the custodial limit of Ref.~\cite{Cort:2013foa} where the five minimization conditions degenerate into only two, 
\begin{align}
&m_H^2+\mu^2=\frac{1}{2}\left(m_3^2 - 3\sqrt{2}v_\Delta (A_\lambda + \lambda\mu_\Delta) -3\lambda(\lambda_3 v_\Delta^2 +(3v_\Delta^2+ v_H^2)\lambda) + 6\sqrt{2} \lambda\mu v_\Delta \right)
\nonumber\\
&m_\Delta^2+ \mu_\Delta^2=\frac{2v_\Delta B_\Delta+ \sqrt{2}(A_{\lambda_3} + 3\lambda_3\mu_\Delta)v_\Delta^2  + 2\lambda_3^2v_\Delta^3   +2v_\Delta\lambda (\lambda_3+3\lambda)v_H^2 +\sqrt{2}(A_\lambda +\lambda(\mu_\Delta-2\mu) )}{-2v_\Delta}
\label{condiciones}
\end{align}

%*********************NeutMass***********************	
%\section{Neutralino Mass Matrix}
%\label{sec:NeutralinoMassMatrix}
%*********************NeutMass***********************
%*********************Bibliography***********************
\bibliography{DM_SCTM}
%*********************Bibliography***********************

\end{document}